\documentclass[preprint,prd,groupedaddress,footinbib,showpacs,rotate]{revtex4}
\usepackage{amssymb,latexsym}
\usepackage{graphicx}
\usepackage{dcolumn}
\begin{document}
%\hfill hep-th/yymmnnn

\title{ T-duality for open strings in the presence of backgrounds and non-commutativity}
\author{C.  A.  Ballon Bayona}
\email{ballon@if.ufrj.br} \affiliation{Instituto de F\'{\i}sica,
Universidade Federal do Rio de Janeiro, Caixa Postal 68528, RJ
21941-972 -- Brazil}

\author{Nelson R. F. Braga}
\email{braga@if.ufrj.br} \affiliation{Instituto de F\'{\i}sica,
Universidade Federal do Rio de Janeiro, Caixa Postal 68528, RJ
21941-972 -- Brazil}
\author{Rafael D'Andrea}
\email{dandrea@if.ufrj.br} \affiliation{Instituto de F\'{\i}sica,
Universidade Federal do Rio de Janeiro, Caixa Postal 68528, RJ
21941-972 -- Brazil}

%\date{\today}

\begin{abstract}
We investigate the effect of T-duality on noncommutativity. 
Starting with open strings ending on a D2-brane wrapped on a $T^2$ torus in the presence of a Kalb Ramond
field, we consider Buscher transformations on the coordinates and background.
We find that the dual system is commutative.
We also study alternative transformations that can preserve noncommutativity.

\end{abstract}


\maketitle

%%%%%%%%%%%%SECAO%%%%%%%%%%%%%%%%
\section{ Introduction }
%%%%%%%%%%%%%%SECAO%%%%%%%%%%%%%%%%
T-duality \cite{Alvarez:1993qi,Alvarez:1994dn,Giveon:1994fu} is one of the most interesting symmetries of string
theory since it relates small scale physics to large scale
physics. When one target space dimension is compact the
strings do not distinguish whether the compactification radius is
$R$ or $\alpha'/R$ and both possible worlds are related by
T-duality. If $d$ space coordinates are compactified in a torus
$T^d$  there are several T-duality transformations associated with
the conformal symmetries of the torus. These transformations act
on the torus metric $g_{ij}$ and also on the winding and momentum numbers 
of the compact coordinates in such a way that the Hamiltonian is preserved.

In the absence of an antisymmetric Kalb-Ramond field $B_{ij}$, some T-duality
transformations can be realized as simple transformations of the
string coordinates.  If this field is turned on, considering the standard T-dualization 
procedure, the metric $g_{ij}$ and coordinates $X^i$  have  a non-trivial transformation. 

In the context of open string theory, the Kalb-Ramond field plays
a crucial role  because may lead to noncommutativity at the string
endpoints \cite{Chu:1998qz,Chu:1999gi,Ardalan:1999av,Seiberg:1999vs}. 
It is interesting to ask whether this noncommutativity
is preserved or not after a T-duality transformation. A very
important discussion of this problem can be found in \cite{Seiberg:1999vs},
where open string background parameters were introduced. 
Another possible approach would be to analyze  the effect of T-duality on the 
boundary conditions. 
D-branes on a noncommutativite torus were studied in \cite{Douglas:1997fm}.
See also \cite{Lizzi:1997em,SheikhJabbari:1999ac,Imamura:2000hs,Maharana:2000fc} for discussions of T-duality 
and noncommutativity. 

The aim of this article is to investigate the effect of  T-duality transformation for open strings in noncommutativity.
We consider as our primal system a D2-brane which wraps a $T^2$-torus in the presence of a constant 
Kalb-Ramond field with spatial components (magnetic field). This is a noncommutative system.  
We perform T-duality by considering background and coordinate transformations of the form proposed by Buscher\cite{Buscher:1987qj}. Analyzing not only the background  but also the boundary conditions we find that the dual system has a {\it commutative} character.  We discuss this result and the fact that the dual target space still allows   noncommutativity, if one starts with different primal systems. Another discussion of noncommutativity and T-duality, for the case of  a background electric field, can be 
found in \cite{DeRisi:2002gt}.

Inspired in the case of zero Kalb-Ramond field, we also consider an alternative transformation consisting on the interchange of the $\tau$ and $\sigma$ derivatives.
When applied to two coordinates, this transformation generates a dual D2-brane with non zero Kalb-Ramond field, preserving noncommutativity.  
We show that this transformation is a symmetry of the Hamiltonian but violates the   condition that winding and momentum modes must be integer for closed strings so it is not a T-duality.   

We begin with a discussion, in section II, of noncommutativity on a $T^2$ torus with Kalb Ramond background. Next, in section III, we make a general discussion of T-duality in terms of transformations involving the
background fields and the winding and momentum numbers  and then zoom in on the case in point: the D2-brane. 
In section IV, we study T-duality for the open string coordinates
considering separately the Buscher dualization of one or two coordinates along the D2-brane.
We also discuss the alternative transformation for these cases. The commutative/noncommutative character of
the dual theories obtained is discussed. 
We follow through with the conclusions.

%%%%%%%%%%%%%%%%%%%%%%%%%SECAO%%%%%%%%%%%%%%%%%%%%%%%%%%%%%%%%%%%%%%%%%%%
\section{D2-brane in a torus with constant Kalb-Ramond field : Open strings noncommutativity}\indent
%%%%%%%%%%%%%%%%%%%%%%%%%SECAO%%%%%%%%%%%%%%%%%%%%%%%%%%%%%%%%%%%%%%%%%%%

Consider a torus $T^2$ formed by two compact angular coordinates 
\begin{equation}
X^{i}\sim X^{i}+2\pi  \, \, \label{winding}
\end{equation}

\noindent with $i=1,2$. Using these angular coordinates the radii of the torus are inserted on the metric. We denote the non-compact coordinates as $X^I$ with $I\neq 1,2$. For simplicity, the only non vanishing component of our Kalb Ramond B field will be $B_{12}=-B_{21}=: \cal B$ where $\cal B$ is a constant.
The toroidal contribution to the  worldsheet action of an open string propagating in the presence of $B_{ij}$ 
can be written as 

\begin{equation}
S=\frac{1}{4\pi } \int d\tau d\sigma [- h^{\alpha\beta}g_{ij}\partial_{\alpha}X^{i}\partial_{\beta}X^{j} + \epsilon^{\alpha \beta}
B_{ij}\partial_{\alpha}X^{i}\partial_{\beta}X^{j}] . \label{stringactionD2brane}
\end{equation}

\noindent where $h_{\alpha \beta } = diag (-,+) $ and the metric $g_{ij}$ is diagonal, with elements $R_i^2/\alpha' $, where $R_{i}$ are the radii of the torus ($i=1,2$).  
The equations of motion are
\begin{equation}
\ddot X^i - X''^i = 0 \, \, \, \, \, \, \, \, i=1,2 \, \, . \label{eqmotion}
\end{equation}

\noindent where $\dot X^i := \partial_\tau X^i $ and $ X'^i := \partial_\sigma X^i \,$.
The action of eq. (\ref{stringactionD2brane}) differs from the free open string case by the Kalb Ramond term which is a surface term that modifies the boundary conditions (BC):

\begin{equation}
\delta X^i(g_{ij}X'^j+B_{ij}\dot X^j)|_{\sigma=0}^{\sigma=\pi}= 0\label{bc}
\end{equation}

\noindent and, consequently, the commutation relations $[X^i(\tau,\sigma),X^j(\tau,\sigma')]$ at the string endpoints, as we shall see. We will choose Dirichlet conditions for the non-compact coordinates $X^I$.

From eq. (\ref{bc}) we see that we have two possible BC for the open string coordinates $X^1$ and $X^2$ at the endpoints: 

i) Dirichlet conditions : $\delta X^i=0$ \, \, , or 

ii) Mixed conditions :  $g_{ij}X'^j+B_{ij}\dot X^j=0$ \, .

The D2-brane correspond to choosing mixed conditions for both $X^1$ and $X^2$ at $\sigma=0, \pi\,$:
\begin{eqnarray}
g_{11}X'^1 +{\cal B}\dot{X^2}&=&0 \nonumber \\
g_{22}X'^2 -{\cal B}\dot{X^1}&=&0. \label{BoundaryD2}
\end{eqnarray}

The solutions for the string coordinates $X^1$ and $X^2$ satisfying the equations of motion and mixed boundary conditions have the following form
\begin{eqnarray}
X^1(\tau,\sigma)&=&x^1+w_1\tau-\frac{\cal B}{g_{11}}w_2\sigma+\frac{i}{n}\alpha_n^-(\tau)\cos n\sigma-\frac{1}{n}\frac{\cal B}{g_{11}}\beta_n^+(\tau)\sin n\sigma\label{expansao de X1}\\
X^2(\tau,\sigma)&=&x^2+w_2\tau+\frac{\cal B}{g_{22}}w_1\sigma+\frac{i}{n}\beta_n^-(\tau)\cos n\sigma+\frac{1}{n}\frac{\cal B}{g_{22}}\alpha_n^+(\tau)\sin n\sigma\label{expansao de X2},
\end{eqnarray}

\noindent where we have introduced the oscillator terms $\alpha_n^{\pm}(\tau):= \alpha_n(\tau)\pm\bar\alpha_n(\tau)$ with $\alpha_n(\tau):= \alpha_ne^{-in\tau}$ and $\bar\alpha_n(\tau):= \bar\alpha_ne^{in\tau}$ and 
the same definitions for  $\beta$. A sum over $n>0$ is implicit. See ref.\cite{Ardalan:1999av,Jing:2005nq} for a similar expansion.

The conjugate momenta are
\begin{eqnarray}
P^1(\tau,\sigma)&=&\frac{1}{2\pi}\frac{{\cal M}}{g_{22}}(w_1+\alpha_n^+\cos n\sigma)\\
P^2(\tau,\sigma)&=&\frac{1}{2\pi}\frac{{\cal M}}{g_{11}}(w_2+\beta_n^+\cos n\sigma)\,,
\end{eqnarray}

\noindent (where ${\cal M}:=g_{11}g_{22}+{\cal B}^2$), and the Hamiltonian can be written as 
\begin{eqnarray}
H(\tau) &=&\frac{\pi}{2}{\cal M}\Bigl[\frac{w_1^2}{g_{22}}+\frac{w_2^2}{g_{11}}+\frac{1}{g_{22}}(\alpha_n\bar\alpha_n+\bar\alpha_n\alpha_n)+\frac{1}{g_{11}}(\beta_n\bar\beta_n+\bar\beta_n\beta_n)\Bigr]\,.
\label{Hamiltoniana}
\end{eqnarray}

We are interested in the commutators $[X^i(\tau,\sigma),X^j(\tau,\sigma')]$, $[X^i(\tau,\sigma),P^j(\tau,\sigma')]$, $[P^i(\tau,\sigma),P^j(\tau,\sigma')]$. It is well known \cite{Chu:1998qz,Chu:1999gi,Ardalan:1999av,Seiberg:1999vs} that the canonical ones are inconsistent with the new boundary conditions (\ref{BoundaryD2}) brought about by the $B_{ij}$ field. The simplest way of obtaining the appropriate commutators is by means of the Heisenberg equations. Comparing the series expansion for 
the commutator $[X^i(\tau,\sigma),H(\tau)]$ with the expansion of $\dot{X}^i(\tau,\sigma)$, we obtain the commutation relations for the modes (see appendix) 
\begin{eqnarray}
[x^1,w_1]=\frac{g_{22}}{\pi{\cal M}} \,\, , \, \, [x^2,w_2]=\frac{g_{11}}{\pi{\cal M}} \, \, ,  \nonumber \\
\left[ \alpha_n ,\bar\alpha_m \right] = -i n \frac{g_{22}}{\pi{\cal M}}\delta_{mn} \, \, , \, \, \left[\beta_n,\bar\beta_m\right]=-in\frac{g_{11}}{\pi{\cal M}}\delta_{mn} \, .
\end{eqnarray}

\noindent which we then use to arrive at the desired commutators

\begin{equation}
[P^i(\sigma),P^j(\sigma')]=0\label{primalmetodoexpansaoPP}
\end{equation}
\begin{equation}
[X^i(\sigma),P^j(\sigma')]=\delta^{ij}\delta_N(\sigma-\sigma') := \frac{1}{\pi}(1+2\cos n\sigma\cos n\sigma')\label{primalmetodoexpansaoXP}
\end{equation}
\begin{equation}
[X^i(\sigma),X^j(\sigma')]=\left\{\begin{array}{rcl}0,&0<\sigma,\sigma'<\pi&\\-\frac{B_{ij}}{{\cal M}},&\;\sigma=\sigma'=0&\\\frac{B_{ij}}{{\cal M}},&\sigma=\sigma'=\pi&\end{array}\right. \,.
\label{primalmetodoexpansaoXX}
\end{equation}

So in the end we are left with a noncommutative theory. Notice that for $B_{ij}=0$ we recover the canonical commutators.

%%%%%%%%%%%%%%SECAO%%%%%%%%%%%%%%%%
\section{T-duality}
%%%%%%%%%%%%%%SECAO%%%%%%%%%%%%%%%%

T-duality appears as a symmetry of closed strings when $d$ coordinates of the spacetime are compactified on a torus $T^d$ :
\begin{equation}
X^{i}\sim X^{i}+2\pi m^{i} \, \, \label{winding}
\end{equation}

\noindent with $m^{i}$ integer numbers ($i=1,..,d$). The radii of the torus are included in the metric. We consider in this article the case $d=2$.

The toroidal contribution to the worldsheet action of a closed string propagating in the presence of a constant Kalb-Ramond field is the same as eq. (\ref{stringactionD2brane}). The difference is that instead of boundary conditions we have periodicity of the closed string coordinates 

\begin{equation}
X^{i} (\tau,\sigma+2\pi )\,=\,  X^{i} (\tau,\sigma)  +2\pi m^{i} \, \, \label{winding}\,.
\end{equation}

The conjugate momenta are
\begin{equation}
P_ i =\frac{1}{2\pi}\big(g_{ij}\dot X^{j} + B_{ij}X'^{j}\big)\,.
\end{equation}

The periodicity of $X^i$ leads to a discretization of the center of mass momenta
\begin{equation}
p_i \,=\, \int_0^{2\pi} d\sigma P_ i \,=\, n_i \,,
\label{momentumnumber}
\end{equation}

\noindent with $n_i \,\epsilon \,  Z \!\!\!\! Z $.

The Hamiltonian corresponding to action (\ref{stringactionD2brane}) can be written as\cite{Giveon:1994fu}
\begin{eqnarray}
H &=& \frac{1}{4 \pi} \int d\sigma \Big[ (2\pi)^2 P_i g^{ij}P_j +
X'^i ( g - B g^{-1} B )_{ij} X'^j+4\pi X'^i  B_{ik} g^{kj} P_j   \Big]  \nonumber\\
&=& \frac{1}{4 \pi} \int d\sigma \big( P_L^2 \,+\, P_R^2 \,\big)\,,
\label{Hamiltonian01}
\end{eqnarray}

\noindent where $P_{L}^2 = P_{L a} P_{L a}$ and $P_{R}^2 = P_{R a} P_{R a}$ with
\begin{eqnarray}
P_{L a} &=& \frac{1}{\sqrt{2}} [ 2\pi P_i + (g - B)_{ij}
X'^j ] e^{\ast \,i}_a =\frac{1}{\sqrt{2}}
g_{ij} ( \dot X^j + X'^j)
e^{\ast \,i}_a \nonumber\\
P_{R a} &=& \frac{1}{\sqrt{2}} [2\pi P_i - (g + B)_{ij}X'^j  ] e^{\ast \,i}_a \,=\, \frac{1}{\sqrt{2}}g_{ij} ( \dot X^j - X'^j) e^{\ast \,i}_a. \label{momleftright}
\end{eqnarray}

\noindent These momenta correspond to the coordinates ${\overline  X}^a = e^a_{\,\,\,\,i} X^i \,$ with the zweibeins defined by
\begin{equation}
e_i^{\,\,a} e_j^{\,\,a} \,=\, g_{ij} \,\,\,; e_i^{\,\,a} e_a^{\ast \, j } \,=\, \delta^j_i\,\,\,;\,\,
  e_a^{\ast \, i } e_a^{\ast \,j } \,=\,  g^{ij}\,.
\end{equation}

The zero mode part of the momenta are
\begin{eqnarray}
p_{L a} &=& \frac{1}{\sqrt{2}} e^{\ast \,i}_a [ n_i + (g - B)_{ij} m_j ]  \nonumber\\
p_{R a} &=&  \frac{1}{\sqrt{2}} e^{\ast \,i}_a [ n_i - (g + B)_{ij} m_j ] \,,
\end{eqnarray}

\noindent where $n_i$ and $ m_j$ are the integer numbers defined in eqs. (\ref{winding}) and
(\ref{momentumnumber}). It is convenient to write this equation in matricial form
\begin{equation}
\label{pmatricial}
p \,:= \, \pmatrix{ p_{L } \cr p_{R } }\,=\,\frac{1}{\sqrt{2}}
\pmatrix{ e^{\ast } (g-B) & e^{\ast} \cr - e^{\ast} (g + B) & e^{\ast} }
\pmatrix{ m \cr n }\,=:  V Z \,,
\end{equation}

\noindent where $Z = (m_i , n_ j )$ is a 4-vector composed by the winding and momentum numbers of the coordinates $X^1$ and $X^2$ which are integer numbers for closed strings.
Using this equation we can express the Hamiltonian as
\begin{equation}
H \,=\, \frac{1}{2} \Big( p_L^2 \,+\, p_R^2 \,\Big)\,+N + {\tilde N} \,=\, \frac{1}{2} Z^t \,M \, Z
\,+\,N  + {\tilde N}\,,
\label{Hamiltonian}
\end{equation}

\noindent where $ N$ and ${\tilde N} $ are the number operators for the oscillator modes and $M$ is the 4x4 matrix
\begin{equation}
M \,=\, V^t V \,=
\pmatrix{ g - B g^{-1} B & B g^{-1} \cr
-g^{-1} B & g^{-1} }\,.
\end{equation}

The closed string theory has to satisfy the Virasoro constraint
\begin{equation}
L_0 \,-\, {\tilde L}_0 \,= \frac{1}{2} \Big( \alpha_0^a \alpha_0^a - {\tilde \alpha}_0^a {\tilde \alpha}_0^a \Big) + N - {\tilde N} \,=\, -\frac{1}{2} ( p_L^2 - p_R^2 ) + N - {\tilde N} \, =\, 0\,,
\label{Virasoro}
\end{equation}

\noindent  where $ \alpha_0^a $ and $ {\tilde \alpha}_0^a$ are the zero modes of right and left sectors of the coordinates ${\overline X}^a $. Note that $X^i$ are angular coordinates of the torus while ${\overline X}^a$ are the usual string coordinates expressed in string units.
The Virasoro constraint can be written as
\begin{equation}
N - {\tilde N} \,=\,\frac{1}{2} ( p_L^2 - p_R^2 ) \,=\,
\frac{1}{2} Z^t \,J \, \, Z\,,
\label{Virasoro2}
\end{equation}

 \noindent where $J$ is the 4x4 matrix

\begin{equation}
J\,=\, \pmatrix{ 0 & I_{_2} \cr I_{_2} & 0 }\,,
\end{equation}

\noindent with $I_{_2}$ a 2x2 identity matrix. 

T-duality is a transformation of the string state and of the background that preserves the 
Hamiltonian (\ref{Hamiltonian}) and the Virasoro constraint (\ref{Virasoro2}).  This transformation acts on the matrix $M$ as
\begin{equation}
 M \to T M T^t  \,,
\end{equation}

\noindent with
\begin{equation}
T\,=\, \pmatrix{ a & b \cr c & d }\,,
\end{equation}

\noindent where $a,b,c,d$ are 2x2 matrices.
The invariance of the Virasoro constraint corresponds to the condition
\begin{equation}
T J T^t \,=\, J \,.
\end{equation}

The Hamiltonian is preserved if the vector $Z$ transforms as $Z \to (T^t)^{-1} Z $ under T-duality
while the number operators $N $, ${\tilde N}$ remain unchanged.
Note that for closed strings the elements of the vector $Z$ must remain integers after this transformation.
The transformation of the matrix $M$ corresponds to a change in the background $g$ and $B$ that can be expressed as \cite{Giveon:1994fu}
\begin{equation}
E := g + B \to E^{dual} \,=\, g^{dual} + B^{dual} \,=\, (aE + b)\cdot(cE + d)^{-1}\,.
\end{equation}

We are interested in particular cases of T dualities that can be interpreted in terms of dualization 
of the string coordinates.
First we consider a transformation matrix of the following form
\begin{equation}
T_{_{X^i}}\,=\, \pmatrix{ 1 - t_{_{X^i}} & t_{_{X^i}} \cr t_{_{X^i}} & 1 - t_{_{X^i}} }\,,
\label{factdual}
\end{equation}

\noindent with
\begin{equation}
t_{_{X^1}} \,=\, \pmatrix{ 1 & 0 \cr 0 & 0 }\,\,\,\,\,;\,\,\,\,\,\,t_{_{X^2}} \,=\, \pmatrix{ 0 & 0 \cr 0 & 1 }
\end{equation}

The effect of the transformation $T_{_{X^i}}=(T_{_{X^i}})^t=(( T_{_{X^i}})^t)^{-1}\,$ on the vector
$Z$ is simply to interchange the winding number and the momentum number of the corresponding coordinate:
 $ m_i \leftrightarrow n_i $.

We now describe the effect of this transformation on the momenta. Let us choose coordinate $ i = 2$.
In this case we have
\begin{equation}
g^{dual}  \,=\, \pmatrix{ \frac{{\cal M}}{g_{22}} & \frac{{\cal
B}}{g_{22}} \cr \frac{{\cal B}}{g_{22}} & \frac{1}{g_{22}} }
\,\,\,\,\,;\,\,\,\,\,\,B^{dual}  \,=\, \pmatrix{ 0 & 0 \cr 0 & 0 }
\label{mettransf}
\end{equation}

\noindent with ${\cal M} := g_{11}g_{22} + {\cal B}^2=detE$.
The corresponding transformed zweibeins are
\begin{equation}
e^{dual}  \,=\, \pmatrix{ \sqrt{g_{11}} &  \frac{{\cal B}}{\sqrt{g_{22}}} \cr 0 &  \frac{1}{\sqrt{g_{22}}} } \,\,\,\,\,;\,\,\,\,\,\,e^{\ast \,{dual}} \,=\, \pmatrix{ \frac{1}{\sqrt{g_{11}}} &
- \frac{{\cal B}}{\sqrt{g_{11}}} \cr 0 & \sqrt{g_{22}} }
\end{equation}

Using these results and eq. (\ref{pmatricial}) we can find the transformation of the momentum vector
$p = (p_L , p_R ) \,$:
\begin{equation}
\label{ptransffactdual} p^{dual}  \,= \, \pmatrix{ p_{L }^1 \cr
p_{L }^2 \cr p_{R }^1 \cr p_{R }^2  }^{{dual}}\,=\, \pmatrix{ 1 &
0 &0& 0 \cr 0 & 1 & 0 & 0 \cr 0& 0& 1& 0 \cr 0& 0& 0 & -1 } \,
\pmatrix{ p_{L }^1 \cr p_{L }^2 \cr p_{R }^1 \cr p_{R }^2  }
=: U_{_{X^2}}\,p\,\,.
\end{equation}

\noindent  So we see that the T-duality transformation $T_{X^2}$ reverses the sign of the momentum component $p_{R}^2$ while preserving the sign of the other momentum components. We have introduced the matrix $U$ that represents the T-dualization of the momentum vector.

Now let us consider the case of simultaneously interchanging the winding numbers $m_1,m_2$ and the momentum numbers $n_1,n_2$. This is done by the matrix
\begin{equation}
T_{_{X^1X^2}}\,=\, \pmatrix{ 0 & I_2 \cr I_2 & 0 } \label{invdual}
\end{equation}

\noindent (note that $T_{_{X^1X^2}}=T_{_{X^1}}+T_{_{X^2}}$). The transformed fields are
\begin{eqnarray}
g^{dual}=\frac{1}{{\cal M}} \pmatrix{ g_{22} & 0 \cr 0 &
g_{11} } \,\,\,\,\,;\,\,\,\,\,\,B^{dual}=\frac{1}{{\cal M}}
 \pmatrix{ 0 & -{\cal B} \cr {\cal B} & 0 } \, ,
\end{eqnarray}

\noindent with the corresponding zweibeins
\begin{equation}
e^{dual}=\frac{1}{\sqrt{{\cal M}}}\pmatrix{ \sqrt{g_{22}} &
0 \cr 0 &  \sqrt{g_{11}} } \,\,\,\,\,;\,\,\,\,\,\,e^{\ast
\,{dual}} \,=\,\sqrt{{\cal M}} \pmatrix{ \frac{1}{\sqrt{g_{22}}} &
0\cr 0 & \frac{1}{\sqrt{g_{11}}} } \, .
\end{equation}

The transformation (\ref{invdual}) in fact inverts the background matrix $E \to 1/E$. The momentum vector transforms as
\begin{equation}
\label{ptransfinvdual} p^{dual}=\pmatrix{ p_{L }^1 \cr
p_{L }^2 \cr p_{R }^1 \cr p_{R }^2  }^{dual}=
\frac{1}{\sqrt{{\cal M}}}\pmatrix{ \sqrt{g_{11}g_{22}} & {\cal B}
&0& 0 \cr -{\cal B} &  \sqrt{g_{11}g_{22}}& 0 & 0 \cr 0& 0&
-\sqrt{g_{11}g_{22}}& {\cal B} \cr 0& 0& -{\cal B} &
-\sqrt{g_{11}g_{22}} } \, \pmatrix{ p_{L }^1 \cr p_{L }^2 \cr p_{R
}^1 \cr p_{R }^2  } =: U_{_{X^1X^2}} p.
\end{equation}

\noindent This transformation is not as simple as that obtained in eq. (\ref{ptransffactdual}).

In the next section we discuss concrete realizations for the
T-dualities of eq.(\ref{ptransffactdual}) and
eq.(\ref{ptransfinvdual}). Before that it is useful to see how the  
open string parameters defined by Seiberg and Witten change under T-duality.
For open strings ending on D2-brane with mixed boundary conditions:
 $ g_{ij} X^{\prime j} + B_{ij}\dot{X^j} \,=\, 0 \,$,
the coordinate propagator at string endpoints can be decomposed in terms of the 
parameters\cite{Seiberg:1999vs}
\begin{equation}
G^{ij}=\Big(\frac{1}{E}\Big)_{sym}^{ij} \, \, \, \, \, \, , \, \, \, \, \theta^{ij}=
\Big(\frac{1}{E}\Big)_{ant}^{ij} \,,
\end{equation}

\noindent where $ sym (ant) $ means symmetric (anti-symmetric) part. 
The equal time commutator is directly related to the parameter $\theta$:
\begin{equation}
[X^i(\tau,\sigma=0) , X^j(\tau,\sigma'=0)] = -2\pi \theta^{ij} \, .
\end{equation}

So, in principle, studying the transformation of $G^{ij}$ and $\theta^{ij}$ after T-duality we can find out the propagator and commutator of the T-dual world.

The $T_{_{X^2}}$ duality transformed open string parameters are
\begin{equation}
G^{dual} \, =\, \Big(\frac{1}{E^{dual}}\Big)_{sym} \, = \,
\frac{1}{g_{11}}\pmatrix{1 & -{\cal B} \cr -{\cal B} & {\cal M}}
\, \, \, \, , \, \, \, \, \theta^{dual} \, =
\,\Big(\frac{1}{E^{dual}}\Big)_{ant} \, = \, 0 \, ,
\end{equation}

\noindent so we should expect a commutative dual theory. For the $T_{_{X^1X^2}}$ duality we find
\begin{equation}
G^{dual} \, =\, \pmatrix{g_{11} & 0 \cr 0 & g_{22}} \, = \, g \,
\, \, \, , \, \, \, \, \theta^{dual} \, =  \pmatrix{0 & {\cal B}
\cr -{\cal B} & 0} \, = \, B \,\, ,
\end{equation}

\noindent indicating a possible non-commutative dual theory. We will see in the next section that, although the dual $\theta^{ij}$ is different from zero, the $T_{_{X^1X^2}}$ transformation of the open string coordinates
is such that a noncommutative primal system turns into a commutative one.

%%%%%%%%%%%%%%%%%SE큐O%%%%%%%%%%%%%%%%%%%%%%%%%%%%%%%
\section{Duality transformations of open string coordinates}
%%%%%%%%%%%%%%%%%SE큐O%%%%%%%%%%%%%%%%%%%%%%%%%%%%%%%%%

From now on, we take a coordinate-focused approach to T-duality. That is, we will be speaking of ``dualizing" the fields $X^i(\tau,\sigma)$, though still keeping in mind that T-duality is a symmetry of the Hamiltonian, and acts upon the momenta. This will enable us to assess the commutative/noncommutative character of the theories connected by T-duality.

We will discuss here two possible ways of dualizing the string coordinates in the presence of a constant Kalb-Ramond field. The first one is the  T-duality mechanism that consists in introducing a Lagrange multiplier in the action which is eventually identified with the dual coordinate(s). By rewriting the action in terms of the dual coordinate, we get the transformations for the background. This approach was developed 
by Buscher\cite{Buscher:1987qj}.

The other mechanism is an extrapolation of the ${\cal B}=0$ T-duality  prescription:  we simply interchange the $\tau$ and $\sigma$ derivatives of the coordinate(s) $X^i $ to be dualized. 
This corresponds to inverting the sign of the right component of the string coordinates ($X^i_R\to\, -X^i_R$). 
As we shall see, although this alternative mechanism preserves the Hamiltonian and the Virasoro constraint
in the same way as T-duality,  they are not equivalent.

We consider two cases:  

\noindent {\bf A.} Dualizing one coordinate:  $X^2$,

\noindent {\bf B.} Dualizing both coordinates $X^1$ and $ X^2$ . 

\noindent To each case, we apply both mechanisms mentioned above.

We start from a D2-brane with constant Kalb-Ramond field wrapping the torus. This primal system is noncommutative, as discussed in section {\bf II},
as a consequence of the mixed boundary conditions at the string endpoints:
\begin{eqnarray}
g_{11}X'^1 +{\cal B}\dot{X^2}&=&0 \nonumber \\
g_{22}X'^2 -{\cal B}\dot{X^1}&=&0. \label{D2BC}
\end{eqnarray}

Other interesting consequence of these BC is that the "winding" and momentum numbers of the open string coordinates $X^1$ and $X^2$ are now related :
\begin{equation}
m^1 \, =\,-\frac{{\cal B}}{{\cal M}} \, n^2 \, \, \, \, \, \,  m^2 \, =\,\frac{{\cal B}}{{\cal M}}\, n^1 
\end{equation}

\noindent so the primal ``winding"  numbers are not integer numbers, unlike the closed string case.

In the following subsections we will discuss the dualization of open string coordinates, backgrounds and boundary conditions and study its effect on noncommutativity. 

%%%%%%%%%%%%%%%%%%SUBSECAO%%%%%%%%%%%%%%%%%%%%%
\subsection{ DUALIZING ONE COORDINATE}
%%%%%%%%%%%%%%%%%%SUBSECAO%%%%%%%%%%%%%%%%%%%%%

\subsubsection{ T-duality  transformation}

Let us begin defining the worldsheet vector
\begin{equation}
v_{\alpha}^2 := \partial_{\alpha} X^2 \,.\label{worldsheetvectors}
\end{equation}

\noindent where $\alpha = \tau, \sigma \,$.
Action (\ref{stringactionD2brane}) becomes
\begin{equation}
S=\frac{1}{4\pi } \int d\tau d\sigma
[-\sqrt{h}h^{\alpha\beta}(g_{11}\partial_{\alpha}X^1\partial_{\beta}X^1
+ g_{22}v_{\alpha}^2v_{\beta}^2) + 2\epsilon^{\alpha \beta} {\cal
B}\partial_{\alpha}X^{1}v_{\beta}^2] \,.\label{torusactionlagrange}
\end{equation}

Now we add the (vanishing) Lagrange multiplier:
\begin{equation}
S\rightarrow S - \frac{1}{2\pi } \int d\tau d\sigma
\epsilon^{\alpha \beta}\partial_\alpha X^{2}_Sv_{\beta}^2 \,.\label{modifiedaction}
\end{equation}

\noindent If we vary this action with respect to the new coordinate $X^{2}_S$, we recover the primal action
(\ref{stringactionD2brane}) when using eq(\ref{worldsheetvectors}). If, instead, we vary with respect to $v_{\alpha}^2$, we find the following equation of motion :
\begin{equation}
v_{\alpha}^2 = -\frac{1}{g_{22}}\, \epsilon_{\alpha}^{\beta}\,
\partial_{\beta}[X^{2}_S +{\cal B}X^1] \, .
\label{eqmotionv}
\end{equation}

Substituting this equation in  (\ref{modifiedaction}) we find the ``dual'' action :
\begin{equation}
S^S=\frac{1}{4\pi } \int d\tau d\sigma
[- h^{\alpha\beta}g_{ij}^S\partial_{\alpha}X^{i}_S\partial_{\beta}X^{j}_S
+ \epsilon^{\alpha \beta}
B_{ij}^S\partial_{\alpha}X^{i}_S\partial_{\beta}X^{j}_S],
\label{torusdualaction}
\end{equation}

\noindent where the dual fields $g_{ij}^S$ and  $B_{ij}^S$ are precisely the same found in 
 eq. (\ref{mettransf}). Thus we note that this coordinate transformation indeed represents a realization
of the T-duality studied in section {\bf III}. Note that $g_{ij}^S$ is non-diagonal.

Consistency between eqs (\ref{worldsheetvectors}) and
(\ref{eqmotionv}) yields the relations between the dual coordinate
$X^2_S(\tau,\sigma)$ and the primal one:
\begin{eqnarray}
\dot X^2_S&=& g_{22}\, X'^{2}-{\cal B}\,\dot X^1 \nonumber \\
X'^2_S&=& g_{22}\dot X^{2}-{\cal B}X'^1 . \label{crucialrelations}
\end{eqnarray}

In terms of coordinates ${\overline  X}^a = e^a_{\,\,\,\,i} X^i \,$
 these transformations read
\begin{eqnarray}
{\dot {\overline  X^2_S }} \,=\,  \sqrt{g_{22}^S}\dot X^2_S &=& 
{\overline  X}^{'2} \,-\, {\overline {\cal B}} \, {\dot {\overline X^1}}
 \nonumber\\
{\overline  X^{'2}_S}\,=\, \sqrt{g_{22}^S} X^{'2}_S &= & 
{\dot {\overline  X^2 }} \,-\, {\overline {\cal B}} \,{\overline  X}^{'1}\,,
\label{standardt}
\end{eqnarray}

\noindent where $\overline{{\cal B}}:={\cal B}/\sqrt{g_{11} \, g_{22}}$ \,.
Note that in the case ${\cal B} = 0 $ this T-duality transformation corresponds to inter-changing the
$\tau$ and $\sigma$ derivatives for the ${\overline X^2} $ coordinate.

Now we use eqs (\ref{momleftright}), (\ref{mettransf})  and (\ref{crucialrelations}) to check the T-duality transformation of the momentum vector. We find
\begin{equation}
P_{S}\,=\,
{\pmatrix{{P_{S L}^1} \cr {P_{S L}^2} \cr
{P_{S R}^1} \cr {P_{S R}^2}}} =  \pmatrix{ 1 &
0 &0& 0 \cr 0 & 1 & 0 & 0 \cr 0& 0& 1& 0 \cr 0& 0& 0 & -1 } \,
\pmatrix{ P_{L }^1 \cr P_{L }^2 \cr P_{R }^1 \cr P_{R }^2  }
= U_{_{X^2}}\,P\,\,.
 \label{transMOMvector}
\end{equation}

\noindent This result is consistent with the transformation of the zero mode momentum vector given in eq (\ref{ptransffactdual}) and confirms that the Hamiltonian is preserved. 

What about noncommutativity? Using the primal D2-brane boundary
conditions (\ref{D2BC}) and relations (\ref{crucialrelations}) , we
find the dual boundary conditions
\begin{eqnarray}
X^{'1} + \frac{{\cal B}}{{\cal M}}X^{'2}_S \,=\,0 \, , \nonumber \\
\dot{X^2_S} \, =\,0 \, .
\end{eqnarray}

\noindent so that $X_S^2$ satisfies Dirichlet boundary conditions
and the other BC is some kind of rotated Neumann condition. Hence,
our former noncommutativity at the string endpoints is lost, and we are left with a
commutative dual system! This result can be checked by computing the commutators of the string coordinate
operators following a procedure similar to that of section {\bf II} (but with zero Kalb Ramond field
and non-diagonal metric).
Below we will give a geometrical picture of this T-dual system
in terms of a tilted (and non-localized) D1-brane. For an interesting discussion of this system see 
\cite{Zwiebach:2004tj}.

\subsubsection{ Alternative transformation }

Let us now consider the other, more direct means of producing dual coordinates. Define the alternative
dual coordinate  $ X^{2}_A $ by
\begin{eqnarray}
\dot X^{2}_A  := g_{22}X'^{2} \, \, \, , \, \, \, X'^{2}_A := g_{22} \dot
X^{2}. \label{relationsBarton}
\end{eqnarray}

\noindent and
\begin{equation}
g^A := \pmatrix{g_{11} & 0 \cr 0 &\frac{1}{g_{22}} } \label{metricBarton}
\end{equation}

This simple operation leads to the same momentum vector
transformation (\ref{transMOMvector}). So the Hamiltonian (\ref{Hamiltonian01}) is preserved.
In terms of the planar coordinates 
${\overline  X}^a = e^a_{\,\,\,\,i} X^i \,$ these transformations
read
\begin{eqnarray}
{\dot {\overline  X^2_A }}=\sqrt{g_{22}^A}\dot X^2_A ={\overline  X^{'2}} \nonumber \\
{\overline  X^{'2}_A}=\sqrt{g_{22}^A} X^{'2}_A ={\dot {\overline  X^2}}\,.
\end{eqnarray} 

Note that for ${\cal B} = 0$ these transformations are the same as those given in the standard mechanism
(\ref{standardt}).

Using the primal D2-brane boundary conditions and
relations (\ref{relationsBarton}) we find the dual ones 
\begin{eqnarray}
g_{11}X'^1+\frac{{\cal B}}{g_{22}}X'^2_A &=&0 \nonumber\\
\dot X_A^2-{\cal B}\dot X^1&=&0 \label{BartonBC} \, ,
\end{eqnarray}

\noindent for $\sigma =0,\pi$. These BC can be rewritten as 
\begin{eqnarray}
{\overline  X^{'1}}+ \overline{{\cal B}}\,{\overline  X^{'2}_A}=0 \nonumber \\
{\dot {\overline  X^2_A }}-\overline{{\cal B}}\,{\dot {\overline  X^1 }}=0
\label{PlanarBC}
\end{eqnarray}

\noindent 
Therefore, ${\overline  X^{1}}$ and ${\overline  X^{2}_A}$ are nothing more than rotations of plain Neumann and Dirichlet coordinates ${\overline Y^1}$ and ${\overline Y^2}$ defined by 
\begin{equation}
\pmatrix{{\overline Y^1} \cr {\overline Y^2}}:=  (\overline{{\cal M}})^{-1/2} \,\pmatrix{1 & \overline{{\cal B}} \cr -\overline{{\cal B}} & 1 } \pmatrix{{\overline  X^{1}} \cr {\overline  X^{2}_A}}\label{coordinatesY}
\end{equation}

\noindent where $\overline{{\cal M}}:=\frac{{\cal M}}{g_{11}g_{22}}$ \, . The coordinates ${\overline Y^1}$ and ${\overline Y^2}$  are typical of a D1-brane so our results tell us that the dual world consists on a 
non-localized tilted D1-brane (with zero Kalb-Ramond field) that  corresponds to a commutative dual 
system. We can construct an action for $X^2_A$ of the type
\begin{equation}
S^A= \,\frac{1}{4\pi } \int d\tau d\sigma
[-\sqrt{h}h^{\alpha\beta}g^A_{ij}\partial_{\alpha}X_A^{i}\partial_{\beta}X_A^{j}
+ \epsilon^{\alpha \beta}B^A_{ij}\partial_{\alpha}X_A^{i}\partial_{\beta}X_A^{j}] ,
\label{torusactionbarton}
\end{equation}

\noindent with $g^A_{ij}$ given by eq. (\ref{metricBarton}). It is straightforward to show that the dual boundary conditions (\ref{BartonBC}) forces us to
define  $B^A_{ij}=0$ which confirms the fact that the dual theory is commutative.

Finally, we can find a connection between the T-duality transformation and the alternative transformation. Using eqs (\ref{crucialrelations}) and
(\ref{relationsBarton}) we get
\begin{equation}
X^2_S = X_A^2 - {\cal B}X^1  \label{schwarzbarton} \, ,
\end{equation}

\noindent which relates the dual coordinates
$X^2_S$ and $X_A^2$. It is straightforward to show that $X^2_S$ is proportional to the Dirichlet 
coordinate ${\overline Y_2} $ transversal to the tilted D1-brane. Using (\ref{schwarzbarton}) we can see that
the actions $S^A$ and $S^S$ of eqs. (\ref{torusdualaction}) and (\ref{torusactionbarton})
are equivalent. These two mechanisms for dualization of one coordinate seem to lead to the same dual 
world. However, if we calculate the matrix $T_{_{X^2}}^A $

\begin{equation}
T_{_{X^2}}^A \, = \, \pmatrix{ 1 & 0 & 0& - {\cal B}  \cr 0&0&0& 1 \cr 0 & 0 & 1 & 0 \cr 
0 & 1 & {\cal B} & 0 } \ne T_{_{X^2}}^S \,=\, T_{_{X^2}},
\end{equation}

\noindent we see that the winding and momentum numbers transform differently from the standard T-duality.
Although we are considering open strings we note that applying $T_{_{X^2}}^A $ to closed
strings leads in general to non integer momentum and winding numbers. This means that this alternative  transformation  is not a T-duality transformation. 
The particular cases of  ${\cal B}$ having integer values would be exceptions.

%%%%%%%%%%%%%%%%%%SUBSECAO%%%%%%%%%%%%%%%%%%%%%
\subsection{DUALIZING TWO COORDINATES}
%%%%%%%%%%%%%%%%%%SUBSECAO%%%%%%%%%%%%%%%%%%%%%

\subsubsection{ T-duality transformation}

Now we introduce two worldsheet vectors 
\begin{equation}
v^1_\alpha:=\partial_\alpha X^1\hspace{3cm}
v^2_\alpha:=\partial_\alpha X^2.\label{V2}
\end{equation}

Then the action analogous to (\ref{modifiedaction}) is
\begin{eqnarray}S&=&\frac{1}{4\pi}\int d^2\sigma\Big[-\sqrt{-h}h^{\alpha\beta}(g_{11}v^1_\alpha v^1_\beta+g_{22}v^2_\alpha v^2_\beta) +2\epsilon^{\alpha\beta}{\cal B}v^1_\alpha v^2_\beta \nonumber\\
&&-2 X_S^1\epsilon^{\alpha\beta}\partial_\alpha v^1_\beta-2
X_S^2\epsilon^{\alpha\beta}\partial_\alpha v^2_\beta\Big]\,.
\end{eqnarray}

We calculate the equation of motion for each $v^i_{\alpha}$:
\begin{eqnarray}
v^1_\alpha&=&\frac{g_{22}}{{\cal M}}\partial_\rho(\epsilon^\rho_\alpha
X_S^1+\frac{{\cal B}}{g_{22}}\delta^\rho_\alpha X_S^2)\nonumber\\
v^2_\alpha&=&\frac{g_{11}}{{\cal M}}\partial_\rho(\epsilon^\rho_\alpha
X_S^2-\frac{{\cal B}}{g_{11}}\delta^\rho_\alpha X_S^2)\,,
\label{v22}
\end{eqnarray}

\noindent and substitute back in the action, to obtain the dual
action:
\begin{equation}
S^S=\frac{1}{2}\int d^2\sigma\;\; h^{\alpha\beta}(g^S_{11}\partial
X_S^1\partial_\beta X_S^2+g^S_{22}\partial X_S^2\partial_\beta
X_S^2)+2\epsilon^{\alpha\beta}B^S_{12}\partial_\alpha
X_S^1\partial_\beta X_S^2\,,
\end{equation}

\noindent with $g^S_{11}=\frac{g_{22}}{{\cal M}},\;g^S_{22}=\frac{g_{11}}{{\cal M}},\;
B^S_{12}=-\frac{{\cal B}}{{\cal M}}.$ The background matrix, then, is inverted:
\begin{equation}
E^S=\left(\begin{array}{cc}g^S_{11} & B^S_{12}\\-B^S_{12} & g^S_{22}\end{array}\right) =\frac{1}{{\cal M}}\left(\begin{array}{cc}g_{22} & -{\cal B}\\{\cal B} & g_{11}\end{array}\right) =\frac{1}{E}.
\label{E22}
\end{equation}

From eqs. (\ref{V2}) and (\ref{v22}) we find the relation between primal and dual coordinates
\begin{eqnarray}
\partial_\alpha X_S^1 = \left(g_{11}\epsilon_\alpha^\rho\partial_\rho X^1 + {\cal B}\partial_\alpha X^2\right)\nonumber\\
\partial_\alpha X_S^2 = \left(g_{22}\epsilon_\alpha^\rho\partial_\rho X^2 - 
{\cal B}\partial_\alpha X^1\right)\label{XS} \,.
\end{eqnarray}

Using these relations in the primal boundary conditions (\ref{D2BC}) we find the dual boundary conditions
\begin{eqnarray}
{\dot X}_S^1 &=&  g_{11}X'^1 +{\cal B}\dot{X^2}\,=\, 0 \nonumber \\
{\dot X}_S^2 &=&   g_{22}X'^2 -{\cal B}\dot{X^1}\,= \, 0. \label{D2BC2}
\end{eqnarray}

\noindent Thus both dual coordinates are Dirichlet-type. Consequently the dual system is a D0-brane
and we have commutativity!

Using the dual zweibein 
\begin{equation}
e^*_S =\sqrt{{\cal M}}\left(\begin{array}{cc}\frac{1}{\sqrt{g_{22}}}&0\\0&\frac{1}{\sqrt{g_{11}}}\end{array}\right)\,
\end{equation}

and eqs. (\ref{E22}) and (\ref{XS}) we find the matrix $U_{_{X^1X^2}}^S$ 
that transforms the momentum vector

\begin{equation}
P_S \,=\,
\left(\begin{array}{c}P^1_{SL}\\P^2_{SL}\\P^1_{SR}\\P^2_{SR}\end{array}\right)=\frac{1}{{\cal M}}\left(\begin{array}{cccc}\sqrt{g_{11}g_{22}}&{\cal B}&0&0\\-{\cal B}&\sqrt{g_{11}g_{22}}&0&0\\0&0&-\sqrt{g_{11}g_{22}}&{\cal B}\\0&0&-{\cal B}&-\sqrt{g_{11}g_{22}}\end{array}\right)\left(\begin{array}{c}P^{1}_L\\P^{2}_L\\P^{1}_R\\P^{2}_R\end{array}\right)\,=\, U_{_{X^1X^2}}^S \,P
\end{equation}

\noindent which is the same found in section {\bf III} for the zero mode momentum vector.

\subsubsection{ Alternative transformation }

In this case the dual coordinates $ X_A^i $ are introduced by
\begin{equation}
\dot X_A^i(\tau,\sigma):=g_{ij}{X'}^j(\tau,\sigma)\hspace{3cm}{X'}_A^i(\tau,\sigma):=
g_{ij}\dot X^j(\tau,\sigma)\,,
\end{equation}

\noindent where $g_{ij}$ is the primal diagonal metric. The dual boundary conditions 
\begin{eqnarray}
\frac{1}{g_{11}}{X'}_A^1-\frac{1}{{\cal B}}\dot X_A^2&=&0\nonumber\\
\frac{1}{g_{22}}{X'}_A^2+\frac{1}{{\cal B}}\dot X_A^1&=&0\,,
\end{eqnarray}

\noindent have the same mixed form of those of the primal system. This shows that the dual system is 
a D2-brane with non zero Kalb Ramond field that has a noncommutative behaviour. 
From these BC we figure out the dual background
\begin{equation}
g^A=g^{-1}\hspace{3cm}B^A=\left(\begin{array}{cc}0&-\frac{1}{{\cal B}}\\\frac{1}{{\cal B}}&0\end{array}\right).
\label{metricAA}
\end{equation}

\noindent Note that the zweibein has inverted too. We proceed to
the transformation of the momenta:
\begin{equation}
P_A\,=\, \left(\begin{array}{c}P^1_{AL}\\P^2_{AL}\\P^1_{AR}\\P^2_{AR}\end{array}\right)=\left(\begin{array}{cccc}1&0&0&0\\0&1&0&0\\0&0&-1&0\\0&0&0&-1\end{array}\right)\left(\begin{array}{c}P^{1}_L\\P^{2}_L\\P^{1}_R\\P^{2}_R\end{array}\right)\,=: U_{_{X^1X^2}}^A \,P\,.
\end{equation}

This transformation is clearly a symmetry of the Hamiltonian, and it also preserves $P_L^2-P_R^2$.
 This time, though, the momentum vector transformation $ U_{_{X^1X^2}}^A $ is not
the same as that obtained using T-duality  ($ U_{_{X^1X^2}}^S$). Moreover, the winding-momentum numbers transformation $T_{_{X^1X^2}}^A $ is different too :

\begin{equation}
T_{_{X^1X^2}}^A \, = \, \pmatrix{ 0 & -\frac{1}{{\cal B}} & 0 & 0 \cr \frac{1}{{\cal B}} & 0 & 0 & 0 \cr 1 & 0 & 0 & -{\cal B} \cr 0 & 1 & {\cal B} & 0 } \ne T_{_{X^1X^2}}^S \,=\, T_{_{X^1X^2}},
\end{equation}

\noindent where $ T_{_{X^1X^2}}$ was defined in eq. (\ref{invdual}). 
Again we note that the matrix $T_{_{X^1X^2}}^A $ applied to closed
strings leads in general to non integer momentum and winding numbers. So, as already pointed out in the case of
transforming only one coordinate, the alternative  transformation  is not a T-duality (the ${\cal B} = 1$ case would be an exception).

Regarding commutativity of position operators, while the dual system  of the $X_S^i$ coordinates and $E^S$ background is a commutative one, the dual system of the $X_A^i$ coordinates is noncommutative. 
Indeed, according to (\ref{primalmetodoexpansaoXX}) and (\ref{metricAA}), we have
\begin{equation}
[X^1_A(\tau, 0),X^2_A(\tau, 0)]= - \frac{B_{12}^A}{{\cal M}^A}=\frac{\frac{1}{\cal B}}{\frac{{\cal M}}{ g_{11}g_{22} {\cal B}^2}}=\frac{{\cal B} g_{11}g_{22}}{{\cal M}}=\,-\,g_{11}g_{22}\,[X^1(\tau, 0),X^2(\tau, 0)]\,.
\end{equation}

\noindent That means: the dual commutator is proportional to the primal one.

%%%%%%%%%SE큐O%%%%%%%%%%%%%
\section{Conclusions}
%%%%%%%%%SE큐O%%%%%%%%%%%%%

We have studied the effect of T-duality in noncommutativity for open string coordinates in the presence 
of a Kalb Ramond antisymmetric background. 
We discussed the fact that the transformation of the background (metric and Kalb ramond field) 
concerns only the behavior of the target space. The transformation of a particular system 
living in the target space,  is defined by the transformation of the boundary conditions.  
We considered as our primal system a D2-brane wrapped on a $T^2 $ torus (target space).
This system has mixed boundary conditions and is a two dimensional non-commutative space. 

Considering the T-dualization of just one coordinate we found a commutative dual system.  
This can be understood from the fact that T-duality of one coordinate transforms the original D2-brane 
into a (non-localized) tilted D1-brane. On the other hand, T-duality transforms the target space 
into another $T^2$ torus without Kalb Ramond field.
The alternative transformation applied to one coordinate leads to an equivalent commutative system.

When T-dualizing both coordinates we found a commutative system since the dual boundary conditions are all of Dirichlet type, indicating that the dual system is a (non-localized) D0-brane.
This is a non trivial result since the dual target space is a $T^2 $ torus with non vanishing Kalb Ramond field. It is important to remark that a different system, like a D2 brane, living in this dual target space will be noncommutative. The commutative/noncommutative character of open strings depends not only on the 
target space but also on the boundary conditions, that define a particular D-brane system.
Even in the primal target space the presence of a Kalb Ramond field does not rule out the 
possibility of a commutative system, like a D0 brane. 

On the other hand, the alternative transformation applied to two coordinates leads to a dual system with mixed boundary conditions, corresponding to a non-commutative D2-brane which is not equivalent 
to the D0 system obtained by T-duality.  
We remark that the alternative transformation is not a T-duality since it does not preserve the condition that, for closed strings,  winding and momentum numbers (in the compact directions) are integer numbers.

It may be surprising that non-commutativity is lost for some T-duality transformations,
but we must remember that the T-duality transformation acts only on the compact coordinates 
$X^i$, $ i=1,2$. The non-compact coordinates $X^I, I=3,4,...$ have their commutation relations unchanged. 
Our noncommutative parameter lives on a torus.
This situation differs from the case of non-commutative quantum field theories
formulated in non-compact spaces where the non-commutativity parameter is taken as a physical quantity. 
Since we expect  T-duality transformation to be a symmetry of open-closed string theory, 
the non-commutativity parameter of the compact dimensions should not be a physical observable.

\acknowledgements The authors are partially supported by CLAF, CNPq and FAPERJ.


\begin{thebibliography}{30}

%\cite{Alvarez:1993qi}
\bibitem{Alvarez:1993qi}
  E.~Alvarez, L.~Alvarez-Gaume, J.~L.~F.~Barbon and Y.~Lozano,
  %``Some global aspects of duality in string theory,''
  Nucl.\ Phys.\  B {\bf 415}, 71 (1994)
  [arXiv:hep-th/9309039].
  %%CITATION = NUPHA,B415,71;%%

%\cite{Alvarez:1994dn}
\bibitem{Alvarez:1994dn}
  E.~Alvarez, L.~Alvarez-Gaume and Y.~Lozano,
  %``An introduction to T duality in string theory,''
  Nucl.\ Phys.\ Proc.\ Suppl.\  {\bf 41}, 1 (1995)
  [arXiv:hep-th/9410237].
  %%CITATION = NUPHZ,41,1;%%

%\cite{Giveon:1994fu}
\bibitem{Giveon:1994fu}
  A.~Giveon, M.~Porrati and E.~Rabinovici,
  %``Target space duality in string theory,''
  Phys.\ Rept.\  {\bf 244}, 77 (1994)
  [arXiv:hep-th/9401139].
  %%CITATION = PRPLC,244,77;%%

%\cite{Chu:1998qz}
\bibitem{Chu:1998qz}
  C.~S.~Chu and P.~M.~Ho,
  %``Noncommutative open string and D-brane,''
  Nucl.\ Phys.\  B {\bf 550}, 151 (1999)
  [arXiv:hep-th/9812219].
  %%CITATION = NUPHA,B550,151;%%

%\cite{Chu:1999gi}
\bibitem{Chu:1999gi}
  C.~S.~Chu and P.~M.~Ho,
  %``Constrained quantization of open string in background B field and
  %noncommutative D-brane,''
  Nucl.\ Phys.\  B {\bf 568}, 447 (2000)
  [arXiv:hep-th/9906192].
  %%CITATION = NUPHA,B568,447;%%

%\cite{Ardalan:1999av}
\bibitem{Ardalan:1999av}
  F.~Ardalan, H.~Arfaei and M.~M.~Sheikh-Jabbari,
  %``Dirac quantization of open strings and noncommutativity in branes,''
  Nucl.\ Phys.\  B {\bf 576}, 578 (2000)
  [arXiv:hep-th/9906161].
  %%CITATION = NUPHA,B576,578;%%

%\cite{Seiberg:1999vs}
\bibitem{Seiberg:1999vs}
  N.~Seiberg and E.~Witten,
  %``String theory and noncommutative geometry,''
  JHEP {\bf 9909}, 032 (1999)
  [arXiv:hep-th/9908142].
  %%CITATION = JHEPA,9909,032;%%


%\cite{Douglas:1997fm}
\bibitem{Douglas:1997fm}
  M.~R.~Douglas and C.~M.~Hull,
  %``D-branes and the noncommutative torus,''
  JHEP {\bf 9802}, 008 (1998)
  [arXiv:hep-th/9711165].
  %%CITATION = JHEPA,9802,008;%%

%\cite{Lizzi:1997em}
\bibitem{Lizzi:1997em}
  F.~Lizzi and R.~J.~Szabo,
  %``Target space duality in noncommutative geometry,''
  Phys.\ Rev.\ Lett.\  {\bf 79}, 3581 (1997)
  [arXiv:hep-th/9706107].
  %%CITATION = PRLTA,79,3581;%%

%\cite{SheikhJabbari:1999ac}
\bibitem{SheikhJabbari:1999ac}
  M.~M.~Sheikh-Jabbari,
  %``A note on T-duality, open strings in B-field background and canonical
  %transformations,''
  Phys.\ Lett.\  B {\bf 474}, 292 (2000)
  [arXiv:hep-th/9911203].
  %%CITATION = PHLTA,B474,292;%%

%\cite{Imamura:2000hs}
\bibitem{Imamura:2000hs}
  Y.~Imamura,
  %``T-duality of non-commutative gauge theories,''
  JHEP {\bf 0001}, 039 (2000)
  [arXiv:hep-th/0001105].
  %%CITATION = JHEPA,0001,039;%%

%\cite{Maharana:2000fc}
\bibitem{Maharana:2000fc}
  J.~Maharana and S.~S.~Pal,
  %``Noncommutative open string, D-brane and duality,''
  Phys.\ Lett.\  B {\bf 488}, 410 (2000)
  [arXiv:hep-th/0005113].
  %%CITATION = PHLTA,B488,410;%%

%\cite{Buscher:1987qj}
\bibitem{Buscher:1987qj}
  T.~H.~Buscher,
  %``Path Integral Derivation of Quantum Duality in Nonlinear Sigma Models,''
  Phys.\ Lett.\  B {\bf 201}, 466 (1988).
  %%CITATION = PHLTA,B201,466;%%

%\cite{DeRisi:2002gt}
\bibitem{DeRisi:2002gt}
  G.~De Risi, G.~Grignani and M.~Orselli,
  %``Space / time noncommutativity in string theories without background
  %electric field,''
  JHEP {\bf 0212}, 031 (2002)
  [arXiv:hep-th/0211056].
  %%CITATION = JHEPA,0212,031;%%

%\cite{Jing:2005nq}
\bibitem{Jing:2005nq}
  J.~Jing and Z.~W.~Long,
  %``Open string in the constant B-field background,''
  Phys.\ Rev.\  D {\bf 72}, 126002 (2005).
  %%CITATION = PHRVA,D72,126002;%%

%\cite{Zwiebach:2004tj}
\bibitem{Zwiebach:2004tj}
  B.~Zwiebach,
 ``A first course in string theory,''
%\href{http://www.slac.stanford.edu/spires/find/hep/www?irn=6000347}{SPIRES entry}
{\it  Cambridge, UK: Univ. Pr. (2004) 558 p}
 
\end{thebibliography}
\end{document}